\documentstyle[pra,aps,epsf]{revtex}

\include{epsf} 
\preprint{} 

\begin{document}
\title{Imaging the Near Field}
\date{\today}
\author{S.Anantha Ramakrishna, J.B. Pendry}
\address{The Blackett Laboratory, Imperial College, London SW7 2BZ, UK}
\author{M.C.K. Wiltshire and W.J. Stewart,}
\address{Marconi Caswell Ltd, Caswell, Towcester, Northants NN12 8EQ, UK}
\maketitle

\begin{abstract}
In an earlier paper we introduced the concept of the perfect lens which
focuses both near and far electromagnetic fields, hence attaining perfect
resolution. Here we consider refinements of the original prescription
designed to overcome the limitations of imperfect materials. In particular
we show that a multi-layer stack of positive and negative refractive media
is less sensitive to imperfections. It has the novel property of behaving
like a fibre-optic bundle but one that acts on the near field, not just the
radiative component. The effects of retardation are included and minimized
by making the slabs thinner. Absorption then dominates image resolution in
the near-field. The deleterious effects of absorption in the metal are
reduced for thinner layers.
\end{abstract}



\section{Introduction}

Conventional optics is a highly developed subject, but has limitations of
resolution due to the finite wavelength of light. It has been thought
impossible to obtain images with details finer than this limit. Recently it
has been shown that a `perfect lens' is in principle possible and that
arbitrarily fine details can be resolved in an image provided that the lens
was constructed with sufficient precision. The prescription is simple: take
a slab of material, thickness $d$, and with electrical permittivity and
magnetic permeability given by, 
\begin{equation}
\varepsilon =-1,\quad \quad \quad \mu =-1
\end{equation}
Given that these conditions are realised, the slab will produce an image of
any object with perfect resolution. The key to this remarkable behaviour is
that the refractive index of the slab is, 
\begin{equation}
n=\sqrt{\varepsilon \mu }=-1
\end{equation}
It was Veselago in 1968$_{}$ \cite{veselago} who first realised that
negative values for $\varepsilon ,\mu $ would result in a negative
refractive index and he also pointed out that such a negative refractive
material (NRM) would act as a lens but it took more than 30 years to realise
the concept of negative refractive index at microwave frequencies \cite
{pendryIEEE,pendry96,smith00,smith01}.

\begin{figure}[h]
\epsfxsize=200pt
\begin{center}\mbox{\epsffile{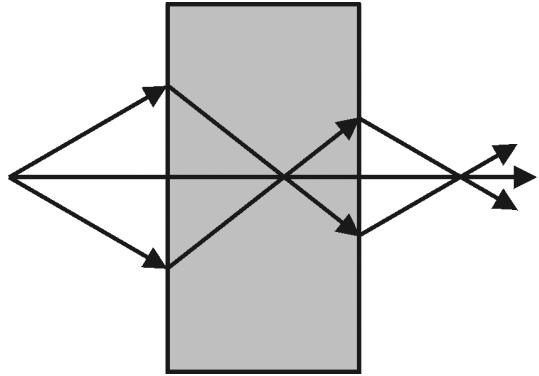}} \end{center}
\caption{A negative refractive index medium bends light to a negative angle
relative to the surface normal. Light formerly diverging from a point
source, is set in reverse and converges back to a point. Released from the
medium the light reaches a focus for a second time outside the medium.}
\end{figure}
It was only in recent times \cite{pendry00} that the lens's remarkable
property of perfect resolution was noted. For the first time there is the
possibility of manipulating the near field to form an image. The physics of
negative refractive index has caught the imagination of the physics
community as evidenced by the publications in the past two years \cite
{smith00,smith01,pendry00,pendry_physworld,ruppin,soukoulis01,tretyakov01a,tretyakov01b,itoh01,sarjmo,solymar,zhang02}

Although the conditions for a perfect lens are simple enough to specify,
realising them is in practice rather difficult. There are two main
obstacles. First the condition of negative values for $\varepsilon ,\mu $
also implies that these quantities depend very sensitively on frequency so
that the ideal condition can only be realised at a single carefully selected
frequency. Second it is very important that absorption, which shows up as a
positive imaginary component of $\varepsilon $ or $\mu $, is kept to a very
small value. Resolution of the lens degrades rapidly with increasing
absorption. It is the objective of this paper to explore how the effects of
absorption can be minimised.

Let us probe a little deeper into the operation of the perfect lens. Any
object is visible because it emits or scatters electromagnetic radiation.
The problem of imaging is concerned with reproducing the electro-magnetic
field distribution of objects in a two dimensional (2-D) plane in the 2-D
image plane. The electromagnetic field in free space emitted or scattered by
a 2-D object (x-y plane) can be conveniently decomposed into the Fourier
components $k_x$ and $k_y$ and polarization defined by $\sigma $: 
\begin{equation}
E(x,y,z;t)=\sum_{k_x,k_y,\sigma }E(k_x,k_y,k_z) 
 \exp [i(k_xx+k_yy+k_zz-\omega
t)],
\end{equation}
where the source is assumed to be monochromatic at frequency $\omega $, $%
k_x^2+k_y^2+k_z^2=\omega ^2/c^2$ and $c$ is the speed of light in free
space. Obviously when we move out of the object plane the amplitude of each
Fourier component changes (note the z-dependence) and the image becomes
blurred. The electromagnetic field consists of a radiative component of
propagating modes with real $k_z,$ and a near-field component of
non-propagating modes with imaginary $k_z$ whose amplitudes decay
exponentially with distance from the source. Provided that $k_z$ is real, $%
\omega ^2/c^2>k_x^2+k_y^2$, it is only the phase that changes with z and a
conventional lens is designed to correct for this phase change. The
evanescent near-field modes are the high-frequency Fourier components
describing the finest details in the object and to restore their amplitudes in
the image plane requires amplification, which is of course beyond the power
of a conventional lens and hence the limitations to resolution.

\begin{figure}[tbp]
\epsfxsize=200pt
\begin{center} \mbox{\epsffile{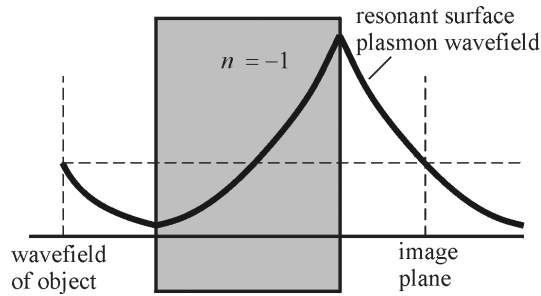}} \end{center}
\caption{The near-field component of an object needs to be amplified before
it can make its contribution to an image. This can be done by resonantly
exciting surface plasmons on the right hand surface. The condition $n = -1$
ensures existence of surface plasmon modes at the operating frequency.}
\end{figure}

Thus the Perfect lens performs the dual function of correcting the phase of
the radiative components as well as amplifying the near-field components
bringing them both together to make a perfect image and thereby eliminating
the diffraction limit on the image resolution. In general the conditions
under which this perfect imaging occurs are : 
\begin{equation}
\varepsilon _{-}=-\varepsilon _{+},~~~~~~~~~~\mu _{-}=-\mu _{+},
\end{equation}
where $\varepsilon _{-}$ and $\mu _{-}$ are the dielectric permittivity and
magnetic permeability of the NRM slab, and $\varepsilon _{+}$ and $\mu _{+}$
are the dielectric permittivity and magnetic permeability of the surrounding
medium respectively.

An important simplification of these conditions can be had in the case that 
{\it all }length scales are much less than the wavelength of light. Under
these circumstances electric and magnetic fields decouple: the P-polarised
component of light becomes mainly electric in nature, and the S-polarised
component mainly magnetic. Therefore in the case of P-polarised light we
need only require that $\varepsilon =-1$, and the value of $\mu $ is almost
irrelevant. This is a welcome relaxation of the requirements especially at
optical frequencies where many materials have a negative values for $%
\varepsilon $, but show no magnetic activity. We shall concentrate our
investigations on these extreme near field conditions and confine our
attentions to P-polarised light.

In Section-2, we investigate the properties of a layered structure
comprising extremely thin slabs of silver and show that layered structures
are less susceptible to the degrading effects of absorption, than are single
element lenses. In section-3, we present some detailed calculations of how
the multilayer lens transmits the individual Fourier components of the image.

\section{The layered perfect lens - an unusual effective medium}

Reference to figure~2 shows that extremely large amplitudes of the electric
field occur within the lens when the near field is being amplified. This is
especially true for the high frequency Fourier components which give the
highest resolution to the image. Unless the lens is very close to the ideal
lossless structure, these large fields will result in dissipation which will
kill the amplifying effect. However there is a way to restructure the lens
to ameliorate the effects of dissipation. We observe that in the ideal
lossless case we can perfectly well divide the lens into separate layers
each one making its contribution to the amplification process (Shamomina et
al have made a similar observation \cite{solymar}  and Zhang et al. have
considered a similar system \cite{zhang02}). Provided that the total length
of vacuum between the object and image is equal to the total length of lens
material, the lens will still work and produce a perfect image. However this
subdivision of the lens makes a big difference to how the lens performs when
it is less than ideal and absorption is present. The point is that by
distribution the amplification, the fields never grow to the extreme values
that they do when the lens is a single slab and therefore the dissipation
will be much less. Figure~3 illustrates this point. 
\begin{figure}[tbp]
\epsfxsize=200pt
\begin{center} \mbox{\epsffile{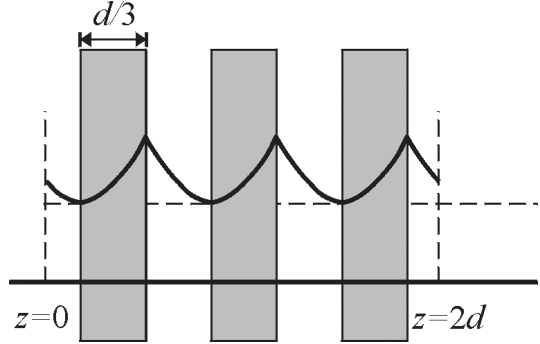}} \end{center}
\caption{Schematic of the field distribution for an incident evanescent wave
on a layered perfect lens, when the original lens is cut into three pieces
placed symmetrically between object and image.}
\end{figure}

First let us estimate the resolution of a lens constituted as a single slab.
According to our original calculations \cite{pendry00} in the near field
limit the transmission coefficient through the lens for each Fourier
component is, 
\begin{equation}
\frac{\exp \left( -2\sqrt{k_x^2+k_y^2}d\right) }{\frac 14\left( \varepsilon
_{-}^{^{\prime \prime }}\right) ^2+\exp \left( -2\sqrt{k_x^2+k_y^2}d\right) }%
,
\end{equation}
where, 
\[
\varepsilon _{-}=\varepsilon _{-}^{^{\prime }}+i\varepsilon _{-}^{^{\prime
\prime }} 
\]
Obviously when, 
\begin{equation}
\sqrt{k_x^2+k_y^2}d<\ln \left[ \frac{\varepsilon _{-}^{^{\prime \prime }}}2
\right]
\end{equation}
the lens' power to amplify begins to fall away. Fourier components of higher
spatial frequency do not contribute and hence the resolution is limited to, 
\begin{equation}
\Delta =\frac{2\pi d}{\ln \left[ \frac{\varepsilon _{-}^{^{\prime \prime }}}2%
\right] }
\end{equation}

\begin{figure}[tbp]
\epsfxsize=200pt
\begin{center} \mbox{\epsffile{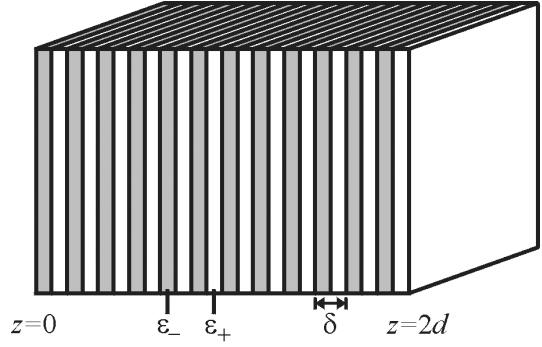}} \end{center}
\caption{In the extreme we can cut the lens into very many thin slices so
that we reduce the effects of absorption as much as possible. In the limit
of infinitesimal slices the ensemble can be treated as an effective medium
with an anisotropic dielectric function.}
\end{figure}

The easiest way to investigate the properties of a layered system is to
recognise that, provided that the slices are thin enough, it will behave as
an effective anisotropic medium whose properties we calculate as follows.
Applying a uniform displacement field, $D$, perpendicular to the slices
gives electric fields of $\varepsilon _0^{-1}\varepsilon _{+}^{-1}D$ and $%
\varepsilon _0^{-1}\varepsilon _{-}^{-1}D$ in the positive dielectric medium
and in the negative material of the lens respectively. Therefore the average
electric field is given by, 
\begin{equation}
\langle E\rangle =\frac 12\left( \varepsilon _0^{-1}\varepsilon
_{+}^{-1}D+\varepsilon _0^{-1}\varepsilon _{-}^{-1}D\right) =\varepsilon
_0^{-1}\varepsilon _z^{-1}D,
\end{equation}
where 
\begin{equation}
\varepsilon _z^{-1}=\frac 12\left( \varepsilon _{+}^{-1}+\varepsilon
_{-}^{-1}\right) ,
\end{equation}
is the effective dielectric function for fields acting along the $z$-axis.
By considering an electric field along the $x$-axis we arrive at 
\begin{equation}
\varepsilon _x=\frac 12\left( \varepsilon _{+}+\varepsilon _{-}\right) ,
\end{equation}
where $\varepsilon _x$ is the effective dielectric function for fields
acting along the $x$-axis. We have assumed for simplicity that the thickness
of each material component is the same, but it is also possible to have
unequal thicknesses. Now under the perfect lens conditions, $\varepsilon
_{-}=-\varepsilon _{+}$, we have 
\begin{equation}
\varepsilon _z\rightarrow \infty ,\quad \varepsilon _x\rightarrow 0
\end{equation}
Thus the stack of alternating extremely thin layers of negative and positive
refractive media in the limiting case of layer thickness going to zero
behaves as a highly anisotropic medium.

Radiation propagates in an anisotropic medium with the following dispersion, 
\begin{equation}
\frac{k_x^2+k_y^2}{\varepsilon _z}+\frac{k_z^2}{\varepsilon _x}=\frac{\omega
^2}{c^2}
\end{equation}
and hence for the perfect lens conditions it is always true that, 
\begin{equation}
k_z=0
\end{equation}
Each Fourier component of the image passes through this unusual medium
without change of phase or attenuation. It is as if the front and back
surfaces of the medium were in immediate contact.

Here we have a close analogy with an optical fibre bundle where each fibre
corresponds to a pixel and copies the amplitude of the object pixels to the
image pixels without attenuation and with the same phase change for each
pixel, preserving optical coherence. Our layered system performs exactly the
same function with the refinement that in principle the pixels are
infinitely small, and the phase change is zero. In figure~5 we illustrate
this point with an equivalent system: an array of infinitely conducting
wires embedded in a medium where $\varepsilon =0$. In the latter case it is
more obvious that an image propagates through the system without distortion.
Indeed in the trivial zero frequency limit the system simply connects object
to image point by point. 
\begin{figure}[tbp]
\epsfxsize=200pt
\begin{center} \mbox{\epsffile{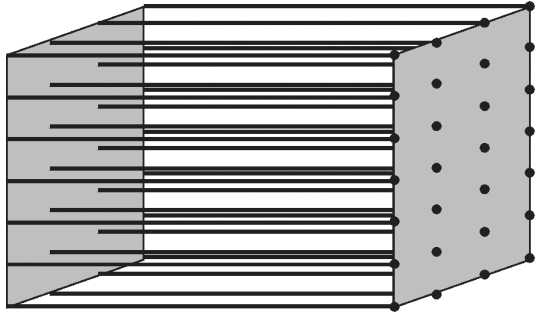}} \end{center}
\caption{An array of very closely spaced infinitely conducting wires,
embedded in a medium where $\varepsilon =0$, behaves in the same manner as
the stack of very thin sheets shown in figure~4.}
\end{figure}

Coming back to our point that the layered system reduces the effect of
absorption, we estimate the transmission for P-polarized light through such
a system in the near field limit as 
\begin{equation}
T_P\simeq \frac 1{\cos \left( \frac i2\varepsilon _{-}^{^{\prime \prime
}}k_x2d\right) +\frac 12\left( \varepsilon _{+}+\varepsilon _{+}^{-1}\right)
\sin \left( \frac i2\varepsilon _{-}^{^{\prime \prime }}k_x2d\right) }.
\end{equation}
Evidently for small values of $k_x$ the transmission coefficient is unity
and these Fourier components contribute perfectly to the image, but for
large values of $k_x$ transmission is reduced. We estimate the resolution
limit to be, 
\begin{equation}
\Delta \simeq \frac{2\pi }{k_{{\rm max}}}=2\pi \left( \frac 12\varepsilon
_{-}^{^{\prime \prime }}2d\right) =2\pi \varepsilon _{-}^{^{\prime \prime
}}d.
\end{equation}
Therefore the smallest detail resolved by the lens decreases linearly with
decreasing absorption ($\varepsilon _{-}^{^{\prime \prime }}$). In contrast
the original single slab of lens had a much slower improvement of
resolution, being inversely as $\ln \left[ \varepsilon _{-}^{^{\prime \prime
}}\right] $. Thus it appears to be a case of {\it two lenses are better than
one but many lenses are the best of all}.

\section{Image Simulations for a multilayer stack}

In the previous section we gave some qualitative arguments as to the
properties of metal-dielectric multilayer stacks and is clear that for
P-polarized light in the quasi-static limit this structure would behave as a
near-perfect `fibre optic bundle'. In the electrostatic (magnetostatic)
limit of large $k_x\sim k_z\sim q_z$, there is no effect of changing $\mu $($%
\varepsilon $) for the P(S)-polarization. The deviation from the
quasi-static limit caused by the non-zero frequency of the electromagnetic
wave would, however, not allow this decoupling. When the effects of
retardation are included, a mismatch in the $\varepsilon $ and $\mu $ from
the perfect-lens conditions would always limit the image resolution and also
leads to large transmission resonances associated with the excitation of
coupled surface modes that could introduce artifacts into the image\cite
{smithprep}. For the negative dielectric (silver) lens, the magnetic
permeability $\mu =1$ everywhere, and this is a large deviation from the
perfect lens conditions. The dispersion of these coupled slab plasmon
polaritons and their effects on the image transfer has been extensively
studied in Ref.~\cite{sarjmo}.

Essentially, for a single slab of negative dielectric material which
satisfies the conditions for the existence of a surface plasmon on both the
interfaces, the two surface plasmon states hybridise to give an
antisymmetric and a symmetric state, whose frequencies are detuned away from
that of a single uncoupled surface state. The transmission as a function of
the transverse wave-vector remains reasonably close to unity up to the
resonant wave-vector for the coupled plasmon state, after which it decays
exponentially with larger wavevectors. The secret for better image
resolution is to obtain a flat transmission coefficient for as large a range
of wave vectors as possible. This is possible by using a thinner slab in
which case the transmission resonance corresponding to a coupled slab mode
occurs at a much larger $k_x$. For the transfer of the image over useful
distances, we would then have to resort to a layered system of very thin
slabs of alternating positive and negative media.

Let us now consider a layered system consisting of thin slabs of silver
(negative dielectric constant $\varepsilon _{-}$) and any other positive
dielectric medium ($\varepsilon _{+}$). Since the dielectric constant of
silver is dispersive\footnote{%
An empirical form for the dielectric constant of silver in the visible
region of the spectrum is $\varepsilon_{-}(\omega) = 5.7 -
9.0^{2}(\hbar\omega)^{-2} + i 0.4$, ($\hbar \omega$ in eV). The imaginary
part can be taken to be reasonably constant in this frequency range.}, we
can choose the frequency ($\omega $) of the electromagnetic radiation so as
to satisfy the perfect lens condition at the interfaces between the media ($%
\varepsilon _{-}(\omega )=-\varepsilon _{+}(\omega )$). We use the transfer
matrix method\cite{bornwolf} to compute the transmission through the layered
medium as a function of the transverse wave-vector at a frequency at which
the perfect lens condition is satisfied. We will denote by $N$ the number of
slabs with negative dielectric constant in the alternating structure, each
period consisting of a negative and positive slab as shown in figure~4. Now
the total length of the system is $2d=N\delta $ where $\delta $ is the
period of the multilayer stack (the negative and positive slabs being of
equal thickness of $\delta /2$). Note that the total thicknesses of positive
and negative dielectric media between the object plane to the image plane
are also equal.

\begin{figure}[tbp]
\epsfxsize=200pt
\begin{center} \mbox{\epsffile{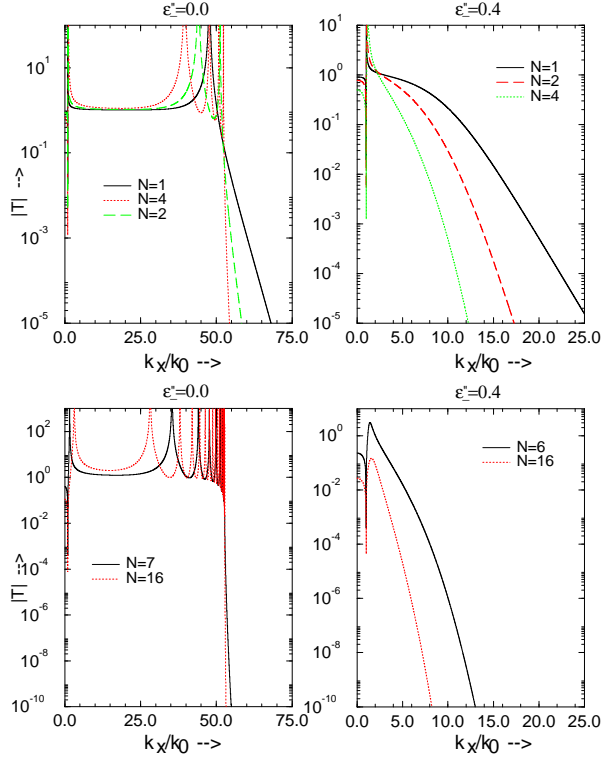}} \end{center}
\caption{The transmission coefficient as a function of $k_x$, for the
metal-dielectric multilayer stack. $\varepsilon _{+}=1.0$, $\varepsilon
_{-}^{\prime }=-\varepsilon _{+}$ and $\delta =$ 20nm. The graphs on the
left are for a hypothetical lossless medium and on the right for silver with 
$\varepsilon _{-}^{\prime \prime }=0.4$ . The layer thickness is kept
constant and the number of layers increased. }
\end{figure}
The transmission across the multilayer system is shown in figure~6, where
the thickness of the individual slabs is kept constant, but the number of
layers is increased, thereby increasing the total length of the system. We
get divergences in the transmission at wave-vectors corresponding to the
coupled plasmon resonances. The number of the resonances increases with the
number of layers, corresponding to the number of surface modes at the
interfaces. For the system with the (hypothetical) lossless negative media,
one notes that as we increase the number of layers, the transmission
coefficient is almost constant and close to unity with increasing $k_x$,
until it passes through the set of resonances and decays exponentially
beyond. The range of $k_x$ for which the transfer function is constant is
independent of the total number of layers and depends only on the thickness
of the individual layers which sets the coupling strength for the plasmon
states at the interfaces. In the presence of absorption in the negative
medium, however, the decay is extremely fast for the system with larger $N$
simply as a consequence of the larger amount of absorptive medium present.
Also note that the absorption removes all the divergences in the
transmission. As noted by us in earlier publications, the absorption is
actually vital in this system to prevent the resonant divergences which
would otherwise create artifacts that dominate the image.

Next we keep the total length of the stack fixed and change the number of
layers. In the lossless case, the range of $k_x$ for which there is
effective amplification of the evanescent waves, simply increases with
reducing layer thickness as can be seen in figure~7. Of course, the number
of transmission resonances which depend on the number of surface states
increases with the number of layers. With absorptive material, however, the
transmission decays faster with $k_x$ for larger $k_x$ in the case of the
thicker slabs (10nm) than in the case of the thinner slabs (5nm). This
reconfirms our analytical result that the effects of absorption would be
less deleterious for the image resolution in the case of thinner layers.
Note that the total amount of absorptive material in this case is the same
in both the cases. 
\begin{figure}[tbp]
\vspace{-70pt}
\epsfxsize=250pt
\begin{center} \mbox{\epsffile{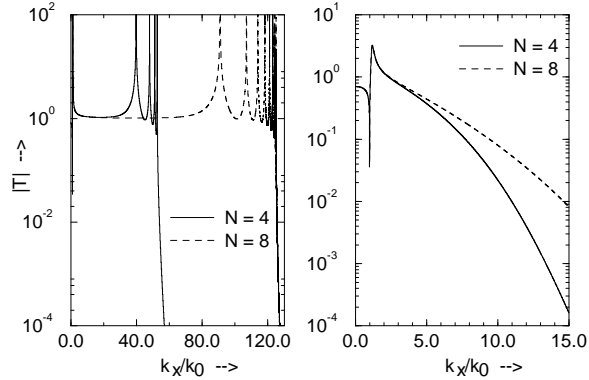}} \end{center}
\caption{The transmission coefficient for two layered stacks of equal total
thickness $2d=$ 80nm, but $\delta/2=$ 10nm $(N=4)$ and $\delta/2=$ 5 nm $%
(N=8)$ in the two cases. The graph on the left is for a hypothetical
lossless medium, and on the right for silver. }
\end{figure}

\begin{figure}[tbp]
\vspace{-70pt}
\epsfxsize=250pt
\begin{center} \mbox{\epsffile{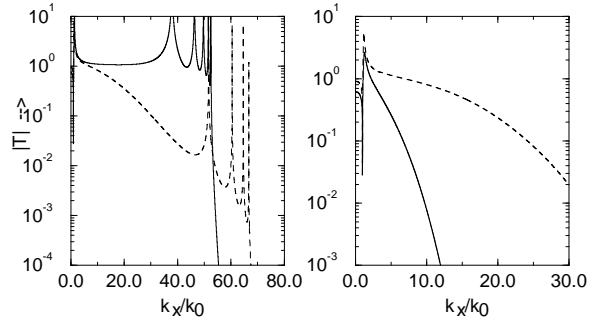}} \end{center}
\caption{The transmission coefficient for two layered stacks of equal total
thickness $2d=$ 100nm and $\delta /2=$ 10 nm. The graph on the left is for a
hypothetical lossless medium, and on the right for silver $\epsilon
_{-}^{\prime \prime }=0.4$. On the right the solid line is for $\epsilon
_{-}^{\prime }=-1$ and the dashed line for $\epsilon _{-}^{\prime }=-12$}
\end{figure}
In any case, the absorption in the negative dielectric (metal) appears to
set the ultimate limit on the image resolution in this case of the layered
medium. We have noted earlier in Ref. \cite{sarjmo} that the effects of
absorption could be minimised by using a large dielectric constant, GaAs say
($\varepsilon _{+}=12$), for the positive medium and tuning to the
appropriate frequency where the perfect lens condition $\varepsilon
_{-}^{^{\prime }}=-\varepsilon _{+}$ is satisfied for the real part of the
dielectric constant $\varepsilon _{-}^{^{\prime }}$ of the metal. In the
case of silver, the imaginary part of the permittivity or the absorption is
reasonably constant ($\sim $0.4) over the frequency range of interest.
Hence, it is immediately seen that the fractional deviation from the perfect
lens condition in the imaginary part is smaller when the real part of the
permittivity is large and hence the amplification of the evanescent waves
becomes more effective. Now we show the transmission obtained across a
multilayer stack where $\varepsilon _{+}=12$ and $\varepsilon _{-}=-12+i0.4$%
, corresponding to alternating slabs of silver and GaAs, in figure~8. We
must first note that the wavelength of light at which the perfect lens
condition for the permittivity of silver is satisfied is different in the
two cases. Using the empirical formula for the dispersion of silver, we
obtain $\varepsilon _{-}^{^{\prime }}=-1$ at 356 nm and $\varepsilon
_{-}^{^{\prime }}=-12$ at 578 nm. In figure~8 for the lossless system, the
transmission resonances appear to occur at higher values of $k_x/k_0$ for
the high index system, but it must realised that $k_0=2\pi /\lambda $ is
smaller in this case and the corresponding image resolution would actually
be lower. However, when we compare the transmission with absorption
included, the beneficial effects of using the larger value of the dielectric
constant become obvious. The transmission coefficient indeed decays much
more slowly with $k_x$ in this case. Also note that we have taken the source
to be in air and the image to be formed inside the high-index dielectric
medium.

Finally, we show in figure~9 the images of two slits of 15nm width and a
peak-to-peak separation of 45 nm obtained by using a single slab of silver
as the lens and a layered medium of alternating layers of silver and a
positive dielectric medium as the lens. The total distance from the object
plane to the image plane in both cases is $2d=$ 80 nm. The images of the
slits in the case of the single slab lens are hardly resolved, whereas the
images of the slits are well separated and clearly resolved in the case of
the layered lens. The enhancement in the image resolution for the layered
lens is obvious from the figure. The bump seen in between the slits is an
artifact due to the fact that the transmission function is not exactly a
constant for all wave-vectors. 
\begin{figure}[tbp]
\vspace{0pt}
\epsfxsize=250pt
\begin{center} \mbox{\epsffile{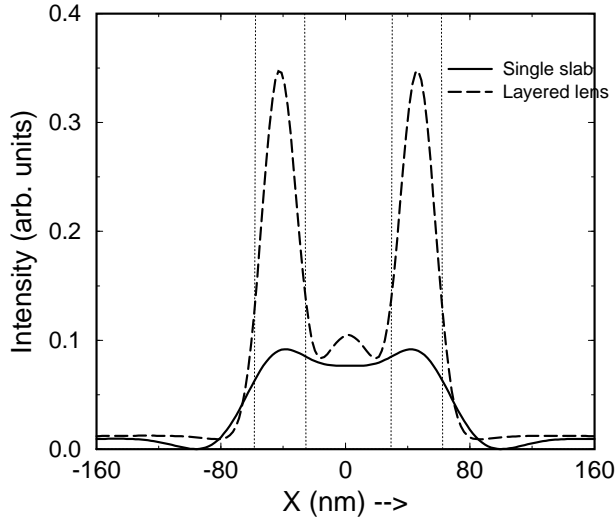}} \end{center}
\caption{The electromagnetic field intensity at the image plane for an
object consisting of two slits of 5nm width and a peak-to-peak separation of
45 nm obtained by (i) using a single slab of silver of thickness 40 nm and,
(ii) a layered stack of alternating positive and negative dielectric layers
of $\delta /2=$ 5 nm layer thicknesses and number of layers is $(N=8)$. $%
\varepsilon _{-}=1+i0.4$ and the object plane to image plane distance is $2d=
$ 80nm in both the cases. }
\end{figure}

\section{Conclusions}

We have elaborated the design of the perfect lens by considering a
multilayer stack and shown that this has advantages over the original
configuration of a single slab of material. In particular the effects of
absorption are much reduced by the division into mutilayers. The limiting
case of infinitesimal multilayers was also considered and shown to be
equivalent to an effective medium through which the image propagates without
distortion as if it were conveyed by an array of very fine infinitely
conducting wires. We went on to make a detailed analysis of how
imperfections in the lens affects the image quality. The effects of
retardation and the coupled slab plasmon resonances can be minimized by
considering very thin layers of 5 to 10 nm thickness. The effects of
absorption then dominate the image transfer, but are less deleterious when
the individual layer thicknesses are smaller. The effects of absorption can
also be minimized by using materials with higher dielectric constants, and
tuning the frequency of the radiation to meet the perfect lens conditions.

\section*{Acknowledgments}

SAR would like to acknowledge the support from DoD/ONR MURI grant
N00014-01-1-0803.

\end{document}